\documentclass[a4paper]{jpconf}
\usepackage{graphicx}
\usepackage{lineno}

\newcommand{\herwig}{\textsc{Herwig}}
\newcommand{\powheg}{\textsc{Powheg}}
\newcommand{\pythia}{\textsc{Pythia}}
\newcommand{\alpgen}{\textsc{Alpgen}}
\newcommand{\madgraph}{\textsc{MadGraph}}
\newcommand{\mcatnlo}{\textsc{MC@NLO}}
\newcommand{\tauola}{\textsc{Tauola}}

\def\antibar#1{\ensuremath{#1\bar{#1}}}

\def\ttbar{\antibar{t}}
\def\pt{\ensuremath{p_{\mathrm{T}}}} 
\def\et{\ensuremath{E_{\mathrm{T}}}} 
\def\met{\ensuremath{E_{\mathrm{T}}^{\mathrm{miss}}}}
\def\het{\ensuremath{H_{\mathrm{T}}^{\mathrm{miss}}}}

\begin{document}
\title{Top quark pair production cross-section results at LHC in final states with tau or no leptons}

\author{Gia Khoriauli \\
On behalf of the ATLAS and CMS Collaborations}

\address{Physik. Inst., Nu\ss{}allee 12, 53115 Bonn, Germany}

\ead{khoriauli@physik.uni-bonn.de}


\begin{abstract}
ATLAS and CMS measured the top quark pair production inclusive cross section using proton-proton collision data collected 
at $7\:$TeV c.m. energy at the LHC during the year $2011$ in several channels. This article presents the measurements of 
the cross section using selections of top quark pair production events with one hadronically decaying $\tau$ lepton or no 
leptons at all.
\end{abstract}


\section{Introduction}
\label{intro}

Top quark pairs, \ttbar, are produced in proton-proton collisions via the strong interaction of colliding partons. 
At $7\;$TeV c.m. energy about $80\%$ of \ttbar~is produced by gluon-gluon interactions. The remaining $20\%$ is produced 
in quark and anti-quark annihilation processes. The theoretical cross section of the inclusive \ttbar~production within 
the Standard Model (SM) is calculated at the approximate NNLO level using the HATHOR tool~\cite{hathor}, and is 
$\sigma^{\rm theory}=167_{-18}^{+17}\;$pb at $\sqrt{s}=7\;$TeV and $m_{\rm top}=172.5\;$GeV.

A top quark decays almost always into a $W$ boson and $b$ quark. Therefore, \ttbar~events are distinguished by the decay 
products of the two $W$ bosons. $W$ bosons can decay into a quark pair or a lepton-neutrino pair. The ATLAS~\cite{atlas} 
and CMS~\cite{cms} experiments measured the \ttbar~production cross section in different final states. These final states 
are affected by different background processes, which makes it necessary to have unique analysis techniques developed for 
each particular measurement. Precise measurements of the cross section in all final states helps to test the SM against 
some new physics phenomena, which could influence on the measurements.  

Details of cross-section measurements of both ATLAS and CMS in three final states are reviewed in this document: 
a) $\tau_{\rm had}$+lepton: one $W$ boson decays into a $\tau$ lepton and a neutrino and the $\tau$ lepton decays 
hadronically. The second $W$ boson decays leptonically including the decay into $\tau$ and neutrino, when the former decays 
leptonically; 
b) $\tau_{\rm had}$+jets: same as in the previous case but the second $W$ boson decays hadronically;
c) fully hadronic: both $W$ bosons decay hadronically. 


\section {Data 2011 and Monte-Carlo (MC) simulation}

Both ATLAS and CMS detectors recorded a large fraction of the total integrated luminosity delivered by the LHC machine during 
the data taking year 2011. The ATLAS and CMS studies, which are presented in this document used up to $4.7\;$fb$^{-1}$ and 
$3.9\;$fb$^{-1}$ of data, respectively. 

The ATLAS signal sample was generated with \mcatnlo~\cite{mcatnlo} and showered and hadronised with \herwig~\cite{herwig}. 
The single top quark background sample was also produced with these generators. In contrast, CMS used 
\madgraph~\cite{madgraph} together with \pythia~\cite{pythia} to generate the signal as well as $W/Z$+jets. The single top 
quark background was generated with \powheg~\cite{powheg} together with \pythia. \pythia~was used by CMS also to generate 
di-bosons and QCD multi-jet samples. The ATLAS studies used di-boson samples generated with \herwig~while the QCD multi-jet 
and $W/Z$+jets background was generated with \alpgen~\cite{alpgen} and \herwig. Both experiments used \tauola~\cite{tauola}
to simulate $\tau$ decays. All top quark samples were generated assuming $m_{\rm top}=172.5\;$GeV. 


\section{Three cross-section measurements}

\subsection{$\tau_{\rm had}$+lepton final state}

The total branching ratio of all hadronic decays of $\tau$ is about $65\%$. The branching ratio of the $\tau_{\rm had}$+lepton final state is about $3.7\%$ of the inclusive \ttbar~production. The background is any process with a $\tau_{\rm had}$+lepton 
or a lepton+jets final state, where one jet is wrongly identified as $\tau_{\rm had}$. The main background processes are 
\ttbar~with different $W$ boson decay modes and $W/Z$+jets.

CMS measured the cross section separately in $t\bar{t} \rightarrow \tau_{\rm had} + e +X$ and 
$t\bar{t} \rightarrow \tau_{\rm had} + \mu +X$ final states using data samples of $2.0\;$fb$^{-1}$ and $2.2\;$fb$^{-1}$ 
respectively~\cite{cmstl}. Signal candidate events were selected using the combined lepton (electron or muon) + dijets + 
\het~trigger, where \het~is the absolute value of the vectorial sum of all jets momenta. The further (offline) selection 
of events required an isolated electron (muon) with a transverse momentum, $\pt > 35\: (30)\;$GeV, and an absolute 
pseudo-rapidity, $|\eta|<2.5\: (2.1)$. A large missing transverse energy, $\met>45\: (40)\;$GeV, and at least two jets 
with $\pt>35\: (30)\;$GeV and $|\eta|<2.4$ were required. At least one of the jets was required to be $b$-tagged. One isolated
$\tau_{\rm had}$ candidate jet was required with $\pt>20\;$GeV, $|\eta|<2.4$ and a charge sign opposite to the selected 
electron (muon). The $\tau_{\rm had}$ algorithm used neutral and charged hadrons in a jet as inputs in order to discriminate 
$\tau_{\rm had}$ from hadronic jets. The background was estimated from MC simulation except for the misidentified 
$\tau_{\rm had}$+lepton background, which was not well modeled in simulation due to the complexity of a precise description 
of numerous detector effects. A data-driven method was used for the estimation of the misidentified $\tau_{\rm had}$+lepton 
background. The method assigns a variable probability to every jet to be misidentified. This probability map, which depends 
on $\pt$, $\eta$ and a width in $\eta \times \phi$ of jets, where $\phi$ is the azimuthal angle around the beam axis, 
was obtained from a data sideband region enriched with QCD multi-jet events. Since one knows the total rate of the processes 
(e.g., $W(\rightarrow l\nu)$+jets), which contribute into the misidentified background one estimates this contribution by 
weighting jets. The cross section was measured using the ``cut and count'' method: the expected number of background events 
was subtracted from the number of selected events in data and the result was normalized to the inverse of the product of 
the signal acceptance (obtained from MC simulation) and the integrated luminosity of data. Systematic uncertainties on 
the expected number of background events, the signal acceptance and the integrated luminosity of data were considered. 
The results are 
$\sigma_{\rm \ttbar} (\ttbar \rightarrow \tau_{\rm had} + e +X) = 136 \pm 23({\rm stat.}) \pm 23({\rm syst.}) \pm 
3({\rm lumi.})$pb and $\sigma_{\rm \ttbar} (\ttbar \rightarrow \tau_{\rm had} + \mu +X)$ \\ 
$ = 147 \pm 18({\rm stat.}) \pm 22({\rm syst.}) \pm 3({\rm lumi.})$pb~\cite{cmstl}.
These results were combined using the Best Linear Unbiased Estimator (BLUE) method~\cite{blue}. The result of the combination 
is $\sigma_{\rm \ttbar} = 143 \pm 14({\rm stat.}) \pm 22({\rm syst.}) \pm 3({\rm lumi.})$pb~\cite{cmstl}.

\begin{figure}[htbp]
\centering
\includegraphics[width=17pc]{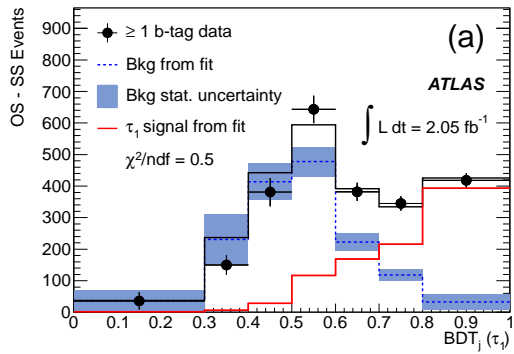}\hspace{2pc}%
\begin{minipage}[b]{17pc}\caption{\label{atlastl1prong} 
${\rm BDT_{j}}$ (OS-SS) templates for $\tau_{\rm had}$+lepton events with the ``1 prong'' $\tau_{\rm had}$ and 
$\ge 1\;b$-jets after the fit to data. The fitted contributions are shown as the light/red (signal), dashed/blue 
(background) and dark/black (total) histograms. Shaded/blue bands are the statistical uncertainty of the background 
template~\cite{atlastl}. }
\end{minipage}
\end{figure}

The ATLAS measurement used $2.05\;$fb$^{-1}$ of data~\cite{atlastl}. Single lepton triggers were used for the event
selection. The offline selection required a single isolated electron (muon) with $\et>25\;$GeV ($\pt>20\;$GeV) and 
$|\eta|<2.47\: (2.5)$ excluding the calorimeter gap regions $1.37<|\eta|<1.52$. Presence of a $\tau_{\rm had}$ candidate 
with $20<\et<100\;$GeV and $|\eta|<2.3$ and not overlapping with the selected lepton, 
$\Delta R({\rm lepton}, \tau_{\rm had}) > 0.4$, was required. In addition, $\met >30\;$GeV and $\met + \sum |\pt|>200\;$GeV, 
where the sum is over the lepton and hadronic jets, requirements were applied. Presence of at least two hadronic jets 
with $\pt>25\;$GeV and $|\eta|<2.5$ and not overlapping with the $\tau_{\rm had}$ candidate, 
$\Delta R({\rm jet}, \tau_{\rm had}) > 0.4$,  was required. Selected events were split in eight categories depending on 
the decay type of the $\tau_{\rm had}$ candidate: with one charged track (``1 prong'') and with three charged tracks 
(``3 prong''); the sign of the isolated lepton and $\tau_{\rm had}$ electric charges: opposite-signed (OS) and same-signed 
(SS); the number of $b$-jets: $\ge 1$ and $0\;b$-jets. A track-based algorithm was used to identify ``1 prong'' and 
``3 prong'' $\tau_{\rm had}$ candidates. A Boosted Decision Tree discriminator, ${\rm BDT}_{e}$ was used to reject 
the background from misidentified electrons. Another ${\rm BDT_{j}}$ discriminator was used for the further selection 
of events. The input templates for ${\rm BDT_{j}}$ were obtained by subtracting the corresponding distributions, OS-SS,
that canceled out the symmetric background in these two selections. Data-driven templates of the fake $\tau_{\rm had}$ 
background due to jets were obtained in the $0\;b$-jet events and corrected for the kinematics of $\ge 1\;b$-jets events 
using weights calculated from MC simulation. Figure \ref{atlastl1prong} shows the ${\rm BDT_{j}}$ output for the ``1 prong'' 
$\tau_{\rm had}$+lepton events with $\ge 1\;b$-jets after the template fit.
Using the template fit of the ${\rm BDT_{j}}$ output for the ``1 prong'' and ``3 prong'' $\tau_{\rm had}$+lepton events 
with $\ge 1\;b$-jets the cross section was measured to be  
$\sigma_{\rm \ttbar} (\ttbar \rightarrow \tau_{\rm had} + e +X) = 186 \pm 15({\rm stat.}) \pm 20({\rm syst.}) \pm 
7({\rm lumi.})$pb and
$\sigma_{\rm \ttbar} (\ttbar \rightarrow \tau_{\rm had} + \mu +X) = 187 \pm 18({\rm stat.}) \pm 20({\rm syst.}) \pm 
7({\rm lumi.})$pb~\cite{atlastl}.
These results were combined using the BLUE method, $\sigma_{\rm \ttbar} = 186 \pm 13(stat.) \pm 20(syst.) \pm 
7({\rm lumi.})$pb~\cite{atlastl}.

\subsection{$\tau_{\rm had}$+jets final state}

The branching ratio of the $\tau_{\rm had}$+jets final state is about $9.8\%$. QCD multi-jet events are one of the main 
backgrounds due to fake $\tau_{\rm had}$ leptons. The combinatorial background, where in a signal event the $\tau_{\rm had}$ 
candidate is misidentified as a hadronic jet, is important, too.

\begin{figure}[htbp]
\centering
\begin{minipage}{17pc}
\includegraphics[width=17pc]{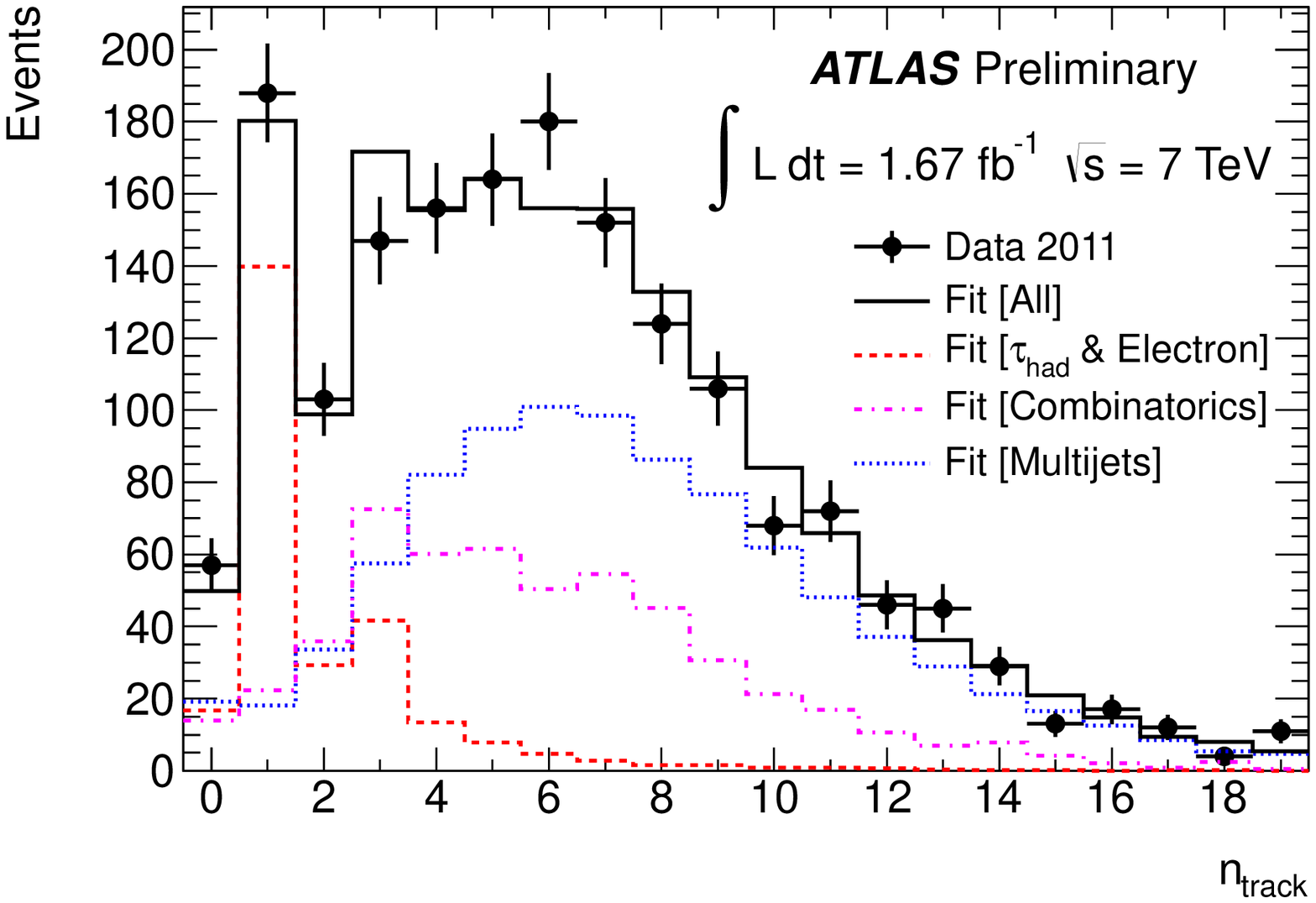}
\caption{\label{atlastjfit} 
The $n_{\rm track}$ distribution for $\tau_{\rm had}$ candidates after all selection cuts. The solid histogram is the result 
of the fit. The red, blue and magenta dashed curves show the fitted contributions from the $\tau_{\rm had}$+electron 
``signal'', the multi-jet and combinatorial backgrounds respectively~\cite{atlastj}.}
\end{minipage}\hspace{2pc}
\begin{minipage}{17pc}
\includegraphics[width=17pc]{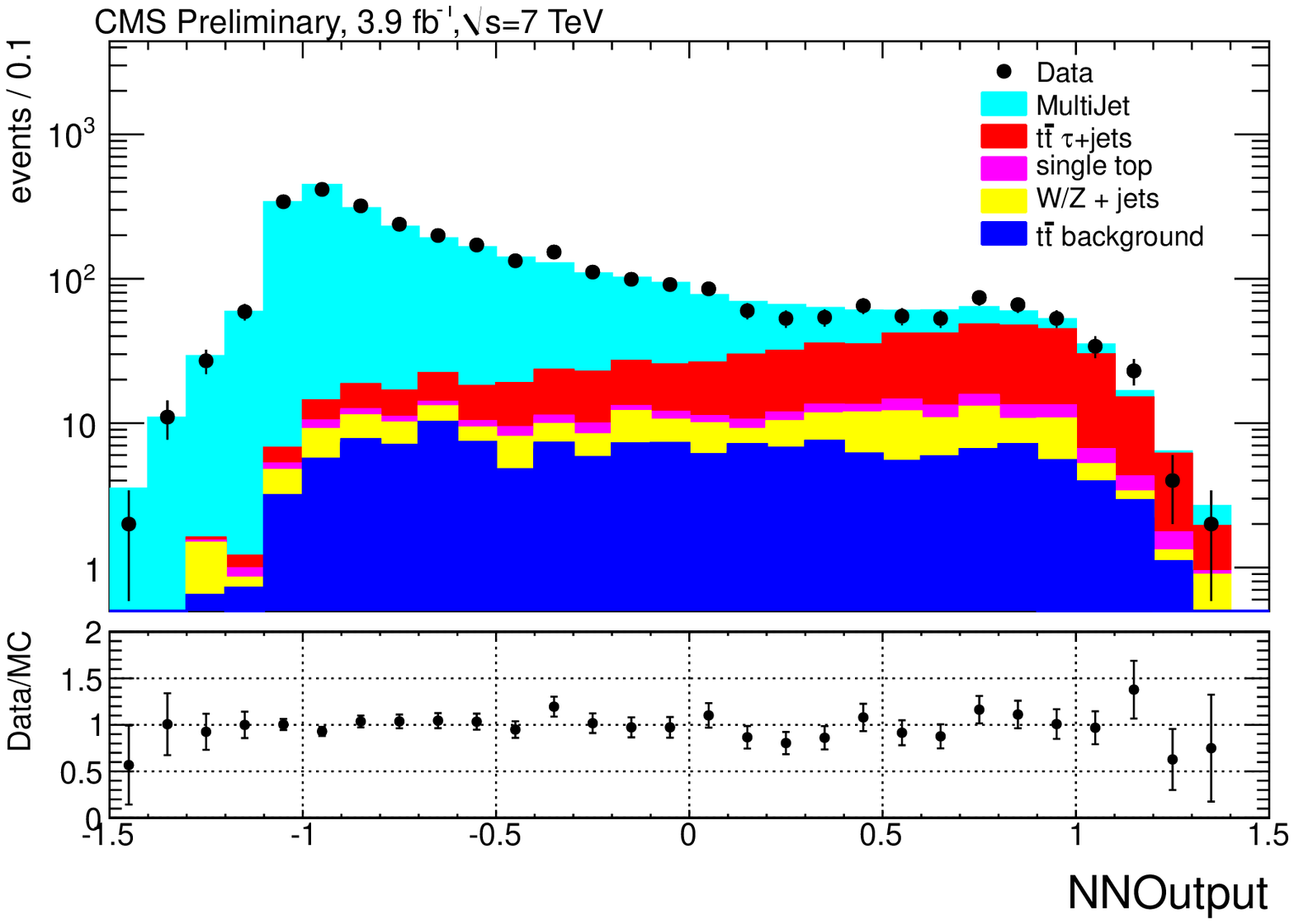}
\caption{\label{cmstjfit}
The distribution of the Neural-Network output after the template fit of the signal (red) and background to data.
The multi-jet template (cyan) is derived from data. The combinatorial \ttbar~(blue), $W/Z$+jets (yellow) and single top 
quark (pink) templates are obtained from MC simulation~\cite{cmstj}.}
\end{minipage}
\end{figure}

The ATLAS measurement used $1.67\;$fb$^{-1}$ of data~\cite{atlastj}. The event selection was done with a multi-jet trigger. 
The offline selection required the presence of at least five jets with $\pt>20\;$GeV, $|\eta|<2.5$ and at least two of 
them tagged as $b$-jets. One of the untagged jets had to be identified as $\tau_{\rm had}$. A veto on isolated leptons 
was applied. To reduce the QCD multi-jet background large $\met$ significance, 
$\met / (0.5 \times \sqrt{\rm GeV} \times \sqrt{\sum{\et}})>8$, was required, where $\sum{\et}$ is the sum of transverse 
energies of all objects contributing to the calculation of $\met$. The $\tau_{\rm had}$ candidate jet was selected by 
excluding other jets using the kinematics and $b$-tagging information. A template fit method was used to measure the signal 
cross section, exploring the number of tracks, $n_{\rm track}$, of the $\tau_{\rm had}$ candidate jet. MC simulation 
predicted a template for misidentified electrons as $\tau_{\rm had}$ candidates similar to the signal template. 
Therefore, the two were joined in a single template. A multi-jet template was derived using a sideband region while 
a template for the combinatorial background was obtained from MC simulation. In the template fit the number of events of 
the combinatorial background was constrained to the number of the signal events since they both originate from 
\ttbar~production. Figure \ref{atlastjfit} shows the fit result. 
Contamination of the fitted signal+electron template by the single top quark and $W$+jets background was subtracted using 
the predictions from MC simulation. The known (MC simulation) fraction of $\tau_{\rm had}$ events, 
$N_{\tau_{\rm had}}/(N_{\tau_{\rm had}} + N_{\rm electron}) = 0.77 \pm 0.03\:({\rm stat.}) \pm 0.03\:({\rm syst.})$, 
was used to estimate the number of $\tau_{\rm had}$ events in the fitted signal+electron template. The result is 
$\sigma_{\rm \ttbar} = 200 \pm 19({\rm stat.}) \pm 43({\rm syst.+lumi.})$pb~\cite{atlastj}.

The CMS measurement used $3.9\;$fb$^{-1}$ of data~\cite{cmstj}. A multi-jet trigger was used for the event selection. 
A jet identified as a $\tau_{\rm had}$ candidate with $\pt>45\;$GeV and $|\eta|<2.3$ was required to match to one of 
the trigger jets. The event selection required the presence of at least three jets with $\pt>45\;$GeV and $|\eta|<2.4$ 
among which exactly one jet is tagged as $b$-jet. Additional jets were counted if $\pt>20\;$GeV. A veto on isolated leptons 
was also applied. Finally, it was required that $\met > 20\;$GeV. A Neural-Network discriminator output was built using 
the MultiLayerPerceptron (MLP) algorithm. The input distributions were derived from MC simulation except for the multi-jet 
background. For the latter, the same signal selection of events was used but with a $0\;b$-jet requirement. Correction 
factors as derived from MC simulation were applied to the distributions in order to take the difference between 
the kinematics of the $0$ and $1\;b$-jet events into account. A result of the template fit of the MLP output is shown 
in Figure \ref{cmstjfit}. The final event selection was done by imposing a cut, ${\rm NNOutput > 0.5}$, on these 
distributions after the fit. The measured cross section is 
$\sigma_{\rm \ttbar} = 156 \pm 12({\rm stat.}) \pm 33({\rm syst.}) \pm 3({\rm lumi.})$pb~\cite{cmstj}.

\subsection{Fully hadronic final state}

The fully hadronic final state has the largest branching ratio, $46\%$. A cross-section measurement suffers from the 
overwhelming QCD multi-jet background. The ATLAS and CMS measurements were based on kinematic fit methods described below. 

\begin{figure}[htbp]
\centering
\begin{minipage}{17pc}
\includegraphics[width=17pc]{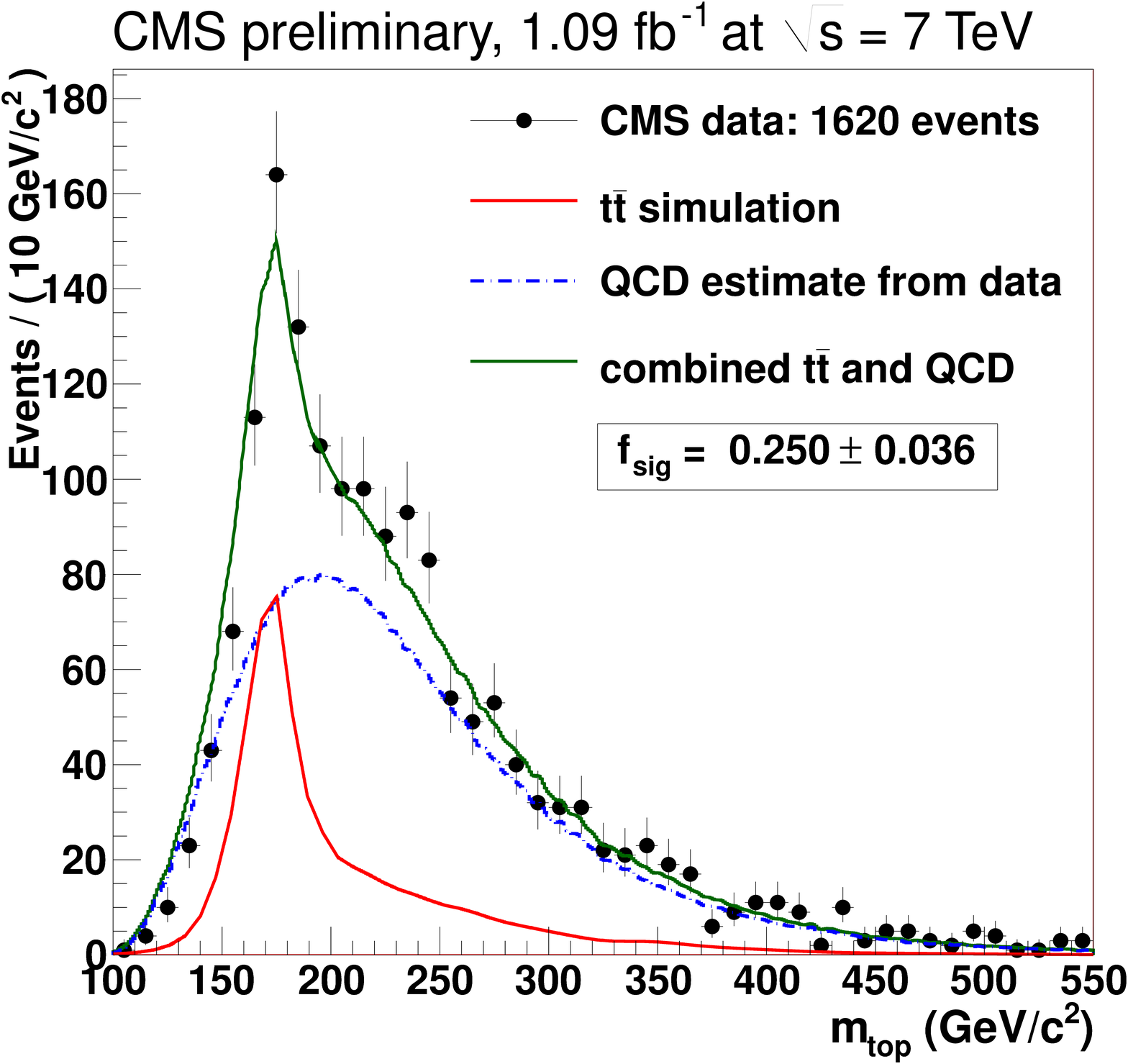}
\caption{\label{cmsfhfit} 
The result of the likelihood fit of the \ttbar~signal (solid red curve) and QCD multi-jet background (dashed blue curve)
templates of the top quark mass (from the kinematic fit) to data. The uncertainty stated on the signal fraction, 
$f_{\rm sig}$, is only statistical~\cite{cmsfh}.}
\end{minipage}\hspace{2pc}
\begin{minipage}{17pc}
\includegraphics[width=17pc]{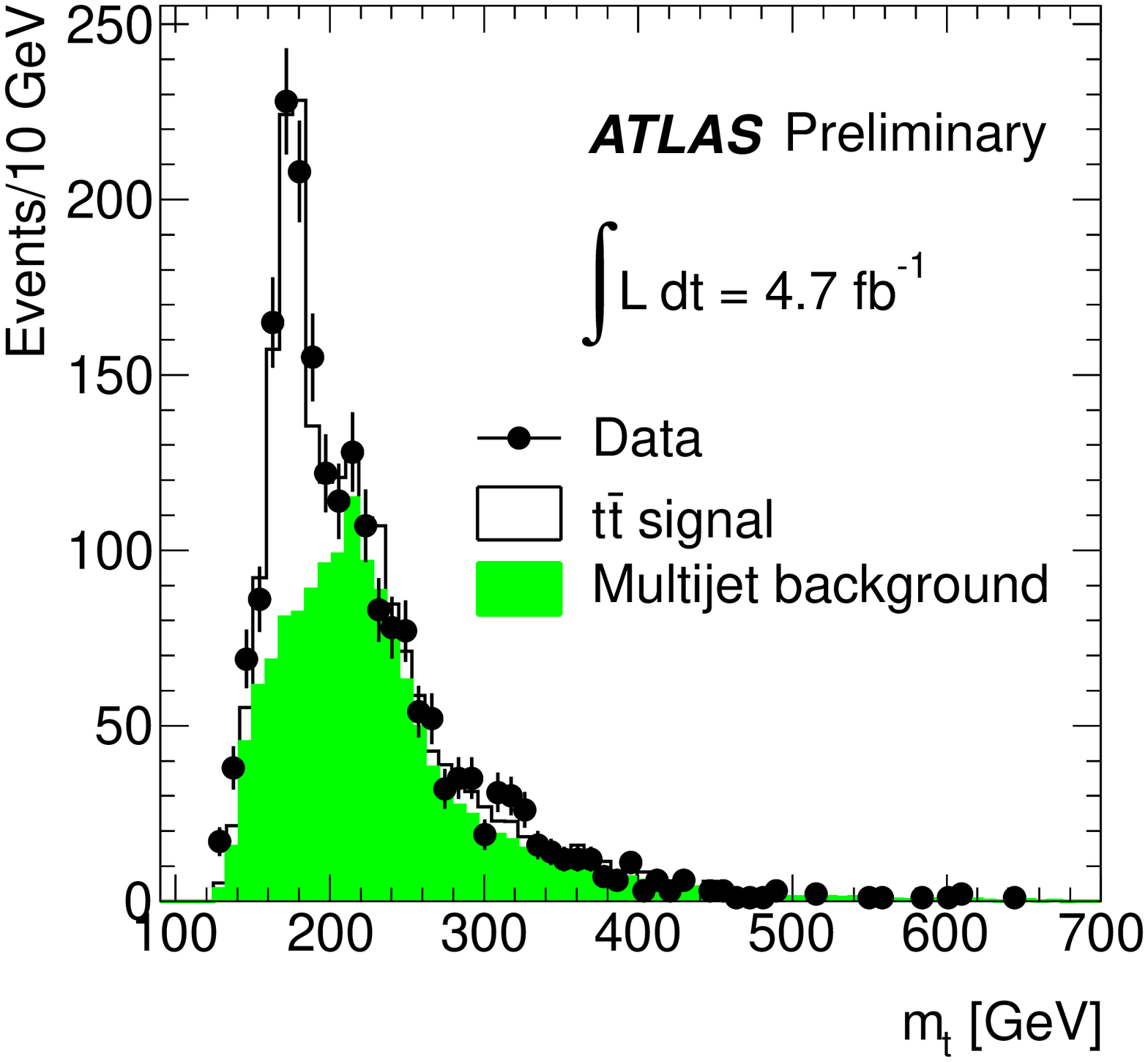}
\caption{\label{atlasfhfit}
The result of the unbinned likelihood fit of the signal (black histogram) and background (green filled histogram) templates
of the top quark mass, $m_{t}$, to data. $m_{t}$ is calculated in the kinematic fit (see text)~\cite{atlasfh}.}
\end{minipage}
\end{figure}

The CMS measurement used  $1.09\;$fb$^{-1}$ of data~\cite{cmsfh}. The event selection was done using a multi-jet trigger.
The offline selection of events required at least six jets with $|\eta|<2.4$ among which four leading jets had to fulfill 
a $\pt>60\;$GeV requirement. The fifth and sixth jets were required to have $\pt>50\;$GeV and $\pt>40\;$GeV, respectively. 
Additional jets with $\pt>30\;$GeV were also counted. At least two jets were required to be $b$-tagged. The jets were used 
to reconstruct the \ttbar~system. All possible permutations of the jets were used to reconstruct two $W$ bosons and two 
top quarks but only the $b$-tagged jets were used as candidates for the true $b$ quark jets. For each permutation 
a least-square fit was done. Gaussian resolutions as derived from MC simulation were used for the energies of the jets.
The true masses of the $W$ bosons were constrained to the known value, $m_{W}=80.4\;$GeV, while both top quark true masses 
were the free parameters of the fit, but constrained to be equal to each other. The permutation  leading to the smallest 
$\chi^{2}$ value was taken and if its probability was greater than $0.01$, the event was selected. The fitted mass in 
the selected events  was used to create templates for a likelihood fit with the signal fraction, $f_{\rm sig}$, as a free 
parameter. The QCD multi-jet template was derived from a sideband selection of events, which was similar to the signal 
selection but requiring no $b$-jets. A two dimensional map ($\eta \times \pt$) of weights, aimed at taking into account 
the difference in kinematics of light jets in events with $0$ and $\ge 2\;b$-jets, was derived using the whole data sample 
(after trigger selection). Since this sample is dominated by events with only light jets, the impact of the true $b$-jets 
on the weights is negligible. The likelihood fit result is shown in Figure \ref{cmsfhfit}. The fitted value of the signal 
fraction was converted to a cross section,
$\sigma_{\rm \ttbar} = 136 \pm 20({\rm stat.}) \pm 40({\rm syst.}) \pm 8({\rm lumi.})$pb~\cite{cmsfh}. 

The ATLAS measurement used $4.7\;$fb$^{-1}$ of data~\cite{atlasfh}. Events were selected with a multi-jet trigger. 
The offline selection required at least six jets with $|\eta|<2.5$ and among them at least five jets with $\pt>55\;$GeV 
and a sixth jet with $\pt>30\;$GeV. At least two out of the leading five jets were required to be $b$-tagged. The jets 
were required to not overlap, $\Delta R({\rm jet}_{i}, {\rm jet}_{k}) > 0.6$, where $i$ and $k$ are the jet indices but 
$i \neq k$. A more stringent cut was applied to the two leading $b$-jets, $\Delta R({\rm bj}_{1}, {\rm bj}_{2}) > 1.2$.
It was required that the \met~significance be $S_{\rm T}=\met / (0.5 \sqrt{\rm GeV} \times \sqrt{H_{\rm T}})<6$, 
where $H_{\rm T}$ is the scalar sum of $\pt$ of all jets. A veto on isolated leptons was applied. All possible permutations
of at least six and up to ten jets were used to construct a likelihood function for every selected event. The free parameters
of the likelihood function were the top quark true mass and the six quark energies. Breit-Wigner constraints were introduced 
in the function for the masses of the two reconstructed $W$ bosons with the fixed true mass and width. Breit-Wigner 
constraints were also used for the reconstructed top quark masses. Both true top quark masses (free parameters) were 
required to be equal to each other. The true width of top quarks was constrained in the fit to vary according to expectation. 
The likelihood function included a product of six transfer functions of quark energies to energies of jets as obtained 
from MC simulation. Also, the likelihood function had factors, which accounted for the  $b$-tagging efficiency and 
the mistagging rate of jets depending on a particular permutation of these jets. These additional factors helped to give 
higher probability to those permutations, where $b$-tagged jets were used as true $b$-jets. An event was selected if 
the permutation with a maximum likelihood value, $L^{\rm max}$, fulfilled the requirement, $L^{\rm max} / \sum L > 0.8$, 
where the sum is over all permutations, and the fitted mass, $m_{t}$, was greater than $125\;$GeV. Using the best 
permutation of jets and fitted values of free parameters, a $\chi^{2}$ value was calculated for the reconstructed 
\ttbar~system in a given event. The event was finally selected if $\chi^{2}<30$. The distributions of the fitted top quark 
mass in the selected events were used in a template fit in order to measure the signal fraction in data. The fit result is 
shown in Figure \ref{atlasfhfit}. The QCD multi-jet template was derived using the same signal selection but requiring no 
$b$-tagged jet. Differences in kinematics of jets in the events with $0$ and $\ge 2\;b$-tagged jets were studied in 
a MC simulation sample of QCD multi-jet production and taken into account as a systematic uncertainty. The measured signal 
fraction was converted into a measurement of the cross section, 
$\sigma_{\rm \ttbar} = 168 \pm 12({\rm stat.}) ^{+60} _{-57}({\rm syst.}) \pm 7({\rm lumi.})$pb~\cite{atlasfh}.


\section{Summary}

The \ttbar~production cross section was measured by the ATLAS and CMS experiments in three different final states 
independently: $\tau_{\rm had}$+lepton, $\tau_{\rm had}$+jets and fully hadronic. The obtained results are consistent with 
the Standard Model prediction for the \ttbar~production cross section.

\ack{I gratefully acknowledge the support of the Deutsche Forschungsgemeinschaft (DFG) through the Emmy-Noether grant 
CR-312/1-2.}


\section*{References}
\begin{thebibliography}{9}
\bibitem{hathor} Aliev M et al 2011 {\it Comput. Phys. Commun.} {\bf 182} 1034
\bibitem{atlas} ATLAS Collaboration 2008 {\it JINST} {\bf 3} S08003
\bibitem{cms} CMS Collaboration 2008 {\it JINST} {\bf 3} S08004
\bibitem{mcatnlo} Frixione S and Webber B R 2002 {\it JHEP} {\bf 06} 029;
			   Frixione S et al 2003 {\it JHEP} {\bf 08} 007;  
			   Frixione S et al 2006 {\it JHEP} {\bf 03} 092
\bibitem{herwig}  Corcella G et al 2001 {\it JHEP} {\bf 01} 010; 
			   Corcella G et al {\it Preprint} hep-ph/0210213
\bibitem{madgraph} Alwall J et al 2007 {\it JHEP} {\bf 09} 028
\bibitem{pythia} Sj\"ostrand T et al 2006 {\it JHEP} {\bf 05} 026
\bibitem{powheg} Frixione S et al 2007 {\it JHEP} {\bf 11} 070
\bibitem{alpgen} Mangano M L et al 2003 {\it JHEP} {\bf 07} 001
\bibitem{tauola} Davidson N et al {\it Preprint} hep-ph/1002.0543
\bibitem{cmstl} CMS Collaboration 2012 {\it Phys. Rev. D} {\bf 85} 112007
\bibitem{blue} Lyons L et al 1988 {\it NIM} A {\bf 270} 110
\bibitem{atlastl} ATLAS Collaboration 2012 {\it Phys. Lett.} B {\bf 717} 89-108 
\bibitem{atlastj} ATLAS Collaboration ATLAS-CONF-2012-032 \verb"http://cdsweb.cern.ch/record/1432198"  
\bibitem{cmstj} CMS Collaboration CMS-PAS-TOP-11-004 \verb"https://cdsweb.cern.ch/record/1446652" 
\bibitem{cmsfh} CMS Collaboration CMS-PAS-TOP-11-007 \verb"http://cdsweb.cern.ch/record/1371755"
\bibitem{atlasfh} ATLAS Collaboration ATLAS-CONF-2012-031 \verb"http://cdsweb.cern.ch/record/1432196"

\end{thebibliography}
\end{document}